# Spontaneous Enhancement of Dzyaloshinskii-Moriya Interaction via Field-Cooling-Induced Interface Engineering in 2D van der Waals Ferromagnetic ternary Tellurides

Shian Xia[1, #], Yan Luo[2, #], Iftikhar Ahmed Malik[1, #], Xinyi Zhou[1], Keying Han[3], Yue Sun[1], Haoyun Lin[1], Hanqing Shi[4], Yingchun Cheng[3], Vanessa Li Zhang[1], Yi Du[4], Sheng Liu[1*], Chao Zhu[2*], Ting Yu[1,5,6*]

[1]School of Physics and Technology, Wuhan University, Wuchang District, Hubei 430072, China.

[2]SEU-FEI Nano-Pico Center, Key Lab of MEMS of Ministry of Education, School of Integrated Circuits, Southeast University, Nanjing 210096, China.

[3]Hebei Key Laboratory of Microstructure Materials Physics, School of Science, Yanshan University, Qinhuangdao, 066004, China.

[4]School of Physics, Beihang University, Haidian District, Beijing 100191, China.

[5]Wuhan Institute of Quantum Technology, Wuhan 430206, China.

[6]Key Laboratory of Artificial Micro- and Nano- structures of Ministry of Education, Wuhan University, Wuhan 430072, China.

\# These authors contributed equally to this work.

\* Correspondence to E-mail: yu.ting@whu.edu.cn; phczhu@seu.edu.cn; liu.sheng@whu.edu.cn

## Abstract

The emergence of two-dimensional (2D) van der Waals (vdW) ferromagnets has opened new avenues for exploring topological spin textures and their applications in next-generation spintronics. Among these materials, $Fe_3GaTe_2$ (FGaT) emerges as a model system due to its room-temperature skyrmion phases, which are stabilized by strong Dzyaloshinskii-Moriya interaction (DMI). However, the atomistic origins of DMI in centrosymmetric vdW lattices remain elusive. Here, we report a spontaneous DMI enhancement mechanism driven by FC in FGaT and its analog $Fe_3GeTe_2$ (FGeT). Combining Raman spectroscopy and scanning transmission electron microscopy (STEM), we have observed the irreversible precipitation of $FeTe_2$ in annealed FGaT. The resulting $FeTe_2$/FGaT heterostructure is considered to break the symmetry and significantly enhance the DMI. Furthermore, similar phenomenon has been observed in the family ferromagnetic material FGeT as well. Additionally, the precipitation of $FeTe_2$ varies significantly with different thicknesses of FGaT, aligning closely with the reported behavior of skyrmions. This discovery provides new insights into the mechanisms behind the origin of the DMI in ternary tellurides, paving the way for advanced spintronic applications.

## Introduction

Since long-range magnetic orders were realized in atomically thin materials, including ferromagnetism (FM, e.g., $CrI_3$ and $Cr_2Ge_2Te_6$[1,2]) and antiferromagnetism (AFM, e.g., $FePS_3$[3,4]), more and more two-dimensional (2D) magnets are being investigated by thinning down van der Waals (vdW) magnetic materials. Plenty of new magnetic properties emerge at the 2D limit, such as layer-dependent magnetism[1,5-7], strong magnetic anisotropy[2,8-10], quantum anomalous Hall effect (QAHE)[11,12] and nonlinear topological magnetic structures[13-15], offering unprecedented opportunities for miniaturized spintronic devices. Among these emerging phenomena, magnetic skyrmions—topologically protected spin textures—have garnered intense interest due to their potential as information carriers in ultrahigh-density memory and logic architectures[15-17].

The formation and stabilization of skyrmions are strongly related to the Dzyaloshinskii-Moriya interaction (DMI), which arises from broken inversion symmetry and induces chiral spin textures[18-20]. Over an extended duration, a proper DMI for generation of skyrmions relies on growth of chiral magnetic materials[14,21] or fabrication of delicately designed heterostructures with alternative ferromagnetic and heavy-metal films[16,22], where cryogenic temperature is still necessary for the realization of skyrmions[23-25].

Recently, room-temperature skyrmions and skyrmion lattice are generated in 2D ferromagnetic material $Fe_3GaTe_2$ (FGaT), whose curie transition temperature is around 350-380 K, by simple field cooling (FC) at moderate temperature[26-33]. Such breakthrough observations herald a transformative era in the development of vdW magnets and topological magnetic orders. However, there persists a fundamental paradox regarding the formation of skyrmions in FGaT: where does the DMI stem from in such a centrosymmetric crystal structure ($P6_3/mmc$). Recent studies have proposed diverse origins for DMI in FGaT, including Fe vacancies[30], atomic displacement[28,29], surface oxidization[34,35] and intercalation of Fe atoms[32]. While these studies propose miscellaneous hypotheses regarding the origin of DMI, a unified understanding of the dominant mechanism has yet to emerge.

Apart from the above observations, an unavoidable fact has been largely neglected. To generate skyrmions in FGaT, thermal treatment during the FC on the sample is unavoidable, and tellurides, especially ternary compounds, are prone to phase transitions or surface reconstruction[36,37]. Therefore, in this study, we investigate the influence of the FC on DMI and skyrmion formation in FGaT and $Fe_3GeTe_2$ (FGeT), revealing the mechanisms of the enhanced DMI in these materials. By combining the studies of Raman spectroscopy and scanning transmission electron microscopy (STEM), we observe the irreversible precipitation of $FeTe_2$ on the surface of FC annealed FGaT. The resulting $FeTe_2$/FGaT heterostructure is considered to be the origin of the enhanced DMI due to the broken inversion symmetry. Furthermore, similar phenomenon has been observed in the family ferromagnetic material FGeT as well. Additionally, the study explores the impact of sample thickness on $FeTe_2$ precipitation and skyrmion formation, indicating $FeTe_2$ precipitation plays a crucial role in enhancing DMI and stabilizing skyrmions. Our findings not only resolve the paradox of DMI emergence in nominally symmetric vdW systems but also establish a universal pathway for engineering topological spin textures through controlled interface engineering in these materials.

## Results

**Emergence of anomalous Raman signatures in FGaT during field cooling**

Pristine FGaT exhibits a vdW layered structure, crystallizing in the *P6₃/mmc* space group, as schematically illustrated in Fig. 1a. Each unit cell consists of a $Fe_3Ga$ slab sandwiched between two Te layers. Within the $Fe_3Ga$ sublayer, Fe atoms occupy two distinct Wyckoff positions, with $Fe_I$ atoms positioned at the top and bottom, and $Fe_{II}$ atoms at the center of the slab[33]. The process of generating stable skyrmions in FGaT through FC is illustrated in Fig. 1. In its pristine state, FGaT exhibits a labyrinth-like domain structure with no observable skyrmion formation, as evidenced by the Magnetic force microscopy (MFM) image shown in Fig. 1b. This observation is consistent with the strong perpendicular magnetic anisotropy (PMA) of material, which inhibits the formation of skyrmions under zero-field conditions[38]. To initiate skyrmion formation, the sample was heated to 393 K—above its Curie temperature ($T_C$)—to ensure that material entered the paramagnetic phase and was fully demagnetized. Subsequently, the sample was cooled back to room temperature in the presence of an external magnetic field (~200 Oe). Following the FC process, stable skyrmions were observed at room temperature by MFM, as demonstrated in Fig. 1c. The skyrmions exhibit an average size of approximately 100–200 nm, in line with previously reported values[30].

A significant challenge during the FC process is the unavoidable increase in temperature, which could potentially alter the intrinsic properties of the sample. Conventional wisdom posits that the moderate temperature rise during field cooling exerts negligible impact on the structural integrity of FGaT. However, the intrinsic thermal sensitivity of ternary tellurides prompts a critical inquiry: could even mild thermal perturbations during FC trigger irreversible structural modifications? To address this, a systematic tracking of the lattice dynamics was conducted through the use of Raman spectroscopy. The Raman spectra, presented in Fig. 1d, reveal notable changes before and after the FC process. The spectrum in yellow corresponds to the pristine FGaT at 303 K, where a broad peak around 150 cm$^{-1}$ is observed, representing the lattice vibrational modes of FGaT [39]. Following FC, two sharp peaks at 121 cm$^{-1}$ and 140 cm$^{-1}$ appear in the Raman spectrum (red line), indicating changes in the FGaT crystal structure. These peaks persist after undergoing twice zero-field cooling (ZFC) processes, suggesting irreversible crystal structural changes.

Thermal decomposition of ternary tellurides has been reported to produce Te or FeTe$_2$ phases under external stimuli such as heating or laser irradiation[36,37,40]. To identify the origin of two sharp Raman peaks that appear after field cooling, we performed phonon spectral simulations, which clearly attribute these peaks to FeTe$_2$ precipitation. Fig. 1e and 1f show the calculated phonon dispersion relations for FGaT and the orthogonal FeTe$_2$, respectively. The calculated phonon modes at Γ point of Brillouin zone, along with their corresponding atomic displacement patterns and Raman shifts are shown in Fig. S2 and Fig. S3. The corresponding Raman spectra derived from these simulations are shown in Fig. 1d as black and gray lines for FGaT and FeTe$_2$, respectively. These theoretical spectra exhibit a high degree of correspondence with the experimental data, suggesting that the two sharp peaks observed in the Raman spectrum are attributable to the lattice vibrations of FeTe$_2$, with the peaks at 121 cm$^{-1}$ and 140 cm$^{-1}$ corresponding to the $A_g$ and $B_{1g}$ Raman modes of FeTe$_2$[41].

To further validate this hypothesis, X-ray photoelectron spectroscopy (XPS) measurements were performed on FGaT samples at various temperatures. Fig. S4a and S4b show the fine spectra of Fe 2p in XPS of FGaT at 20 and 160 °C. At 20 °C, the spectrum shows only Fe$^0$ and Fe$^{3+}$, consistent with the pristine FGaT phase[33]. However, the spectrum at 160°C reveals coexistence of Fe$^0$, Fe$^{2+}$, and Fe$^{3+}$, suggesting the presence of both FGaT and FeTe$_2$ phases, given that the valance states of Fe in FeTe$_2$ is Fe$^{2+}$. As demonstrated in Fig. S4c, the proportion of Fe$^{2+}$ increased gradually with increasing temperature, indicating a temperature-driven precipitation of FeTe$_2$.

Additionally, for samples that exhibited $FeTe_2$ Raman signals after annealing, a re-exfoliation process was employed to remove the surface layers and Raman spectroscopy was subsequently performed on the freshly exfoliated region. The resulting Raman spectra were found to be identical to those observed in the pristine samples, indicating that the majority of the $FeTe_2$ precipitates are formed on the surface of FGaT after annealing (Fig. S5).

**Surface structural reconstruction and symmetry breaking at $FeTe_2$/FGaT interface**

Second harmonic generation (SHG) is exquisitely sensitive to the loss of inversion symmetry in crystalline systems, making it a powerful tool for validating interfacial structural changes. To directly probe the symmetry breaking effects induced by $FeTe_2$ precipitation, we performed SHG measurements in Fig. 1g. The SHG signal is absent in pristine FGaT, which can be attributed to its centrosymmetric crystal structure, belonging to the $D_{6h}$ point group. However, after annealing, a subtle SHG signal emerges in the FGaT sample, suggesting a symmetry-breaking effect due to the structural changes induced by the annealing process. It is noteworthy that the orthogonal $FeTe_2$, which precipitates during the annealing process, belongs to the *Pnnm* space group (No. 58) and is classified within the $D_{2h}$ point group. The crystal structure of $FeTe_2$ is also centrosymmetric, and thus, it would not be expected to exhibit an SHG signal. This observation implies that the SHG signal originates from the interface between the $FeTe_2$ layer and the underlying FGaT. The precipitation of $FeTe_2$ during annealing results in an irreversible alteration of the FGaT lattice in the heterostructure, leading to a breaking of inversion symmetry, which expectedly enhances the DMI crucial for stabilizing topological spin textures such as skyrmions.

To further investigate the surface structural reconstruction in the annealed FGaT at atomic scale, aberration-corrected STEM was employed to examine the crystalline structure of FGaT nanoflakes, both before and after thermal treatment. High-angle annular dark-field (HAADF) cross-sectional STEM images of pristine FGaT along the [1 0 0] (Fig. 2a) and [1 2 0] (Fig. 2b) directions, supported by the bright-field (BF)-

STEM image (Fig. S6), reveal atomic lattice without defects. The crystal structure reveals a Fe$_3$Ga slab is intercalated between two tellurium layers, with a vdW gap separating adjacent Te atoms. No intercalation of iron atoms within the vdW gap is observed. To mitigate ion beam damage during focused ion beam (FIB) sample preparation, a protective graphene capping layer was applied to the surface layer of FGaT (Fig. S7). The pristine edge structure of FGaT slab, as shown in the large-area ADF-STEM and BF-STEM images (Fig. 2c, 2d, Fig. S8), remains unchanged, preserving its pristine atomic arrangement.

Upon annealing at 220°C, atomic rearrangements occur at both the edges of FGaT slab, resulting in the formation of distinct non-layered structures. These changes are evidenced by the appearance of additional diffraction spots in the fast Fourier transform (FFT) patterns that do not correspond to FGaT (Fig. 2e, 2f). The ADF-STEM image in Fig. 2g highlights a vdW gap between the newly formed non-layered structure and the FGaT layers, similar to the original vdW gap observed within the FGaT sublayers. To identify the annealed structure, the atomic arrangement of the non-layered phase was systematically examined along the [1 0 0] and [1 2 0] directions of FGaT. As shown in Fig. 2h, the ADF-STEM image reveals orthogonal FeTe$_2$ (*Pnnm*) along the [0 0 1] zone axis, with the lattice spacing of the (2 0 0) plane measured to be 1.97 Å, as confirmed by the corresponding FFT pattern (Fig. 2i), in excellent agreement with the atomic structure model.

Further analysis of large-scale ADF-STEM images of FeTe$_2$ reveals diffraction spots (green square) along the [3 0 1] zone axis, corresponding to the (1 1 3) plane (Fig. S9a-9c). The atomic arrangements along the [3 0 1] direction show a lack of alternating undulations in the tellurium atoms, unlike those observed along the [0 0 1] direction (Fig. S9d, S9e), as demonstrated by the detailed ADF-STEM images and FFT patterns of FeTe$_2$ (Fig. S10). Additionally, the ADF-STEM image along the [1 0 0] direction of FGaT and its FFT pattern confirm that FeTe$_2$ exists along both the [1 0 0] and [$\bar{1}$ 0 1] zone axes (Fig. 2j, 2k, Fig. S11). Given the overlapping FFT patterns and similar atomic structure (Fig. S11), we propose that FeTe$_2$ forms along both the [1 0 0] and [$\bar{1}$ 0 1] directions, influenced by the biaxial orientation of FeTe$_2$ along the [1 2 0] direction of FGaT. The precipitation of FeTe$_2$ with different crystallographic orientation can be attributed to the thermal annealing process, which represents a non-equilibrium thermodynamic transformation. To verify the elemental composition of the annealed FGaT, energy-dispersive X-ray spectroscopy (EDS) was employed, revealing a uniform distribution of Te, Fe, and Ga in the lower region of FGaT (Fig. 2l–2o). Notably, the edge region (area 1) exhibits a significant Te enrichment, with the corresponding EDS spectra (Fig. 2p) showing an atomic ratio of Fe: Te: Ga = 1: 1.9: 0.28, closely matching the stoichiometry of FeTe$_2$, in perfect agreement with the structural analysis. Furthermore, EDS mapping of area 2 indicates diffusion of Fe atoms into the upper FeTe$_2$ layer, as reflected in the spectra (Fig. S12), which suggest the formation of an amorphous phase with an Fe: Te: Ga ratio of 1: 0.19: 0.02. These findings confirm the precipitation of FeTe$_2$ at the edge of FGaT, establishing the FGaT/FeTe$_2$ heterostructure.

**Universal FeTe$_2$ precipitation and interface-driven spin texture evolution in FGeT**

The spontaneous precipitation of FeTe$_2$ through thermal annealing is not merely a phenomenon isolated to FGaT but represents a universal mechanism within the family of ternary telluride, attributable to the low bond energy of the compounds. Recent studies have shown that FGeT, a well-known family member of FGaT, undergoes thermal decomposition upon annealing, resulting in the precipitation of FeTe$_2$ and Fe$_3$Ge phases through a temperature-induced self-decomposition mechanism[37]. As with FGaT, FeTe$_2$ precipitation exerts a comparable effect on FGeT, as illustrated in Fig. 3. The Raman spectrum of pristine FGeT before annealing (Fig. 3a) exhibits a broad peak near 150 cm$^{-1}$, which is attributed to the characteristic peak of FGeT, as previously reported[42]. Upon annealing, the Raman spectrum of FGeT displays two sharp peaks, consistent with the results observed in FGaT. These sharp peaks correspond to the characteristic vibrations of FeTe$_2$, indicating that the precipitation of FeTe$_2$ also occurs in FGeT after thermal treatment.

To further investigate the atomic structure and chemical composition of the annealed FGeT, graphene-capped cross-sectional ADF-STEM imaging and EDS analysis were conducted (Fig. S13). The pristine cross-sectional FGeT nanosheet exhibits the similar atomic arrangement as FGaT (Fig. S14), and the EDS mapping shows a uniform distribution of Fe, Ge, and Te, with an atomic ratio of Fe: Ge: Te = 3: 1: 2 (Fig. S15). Large-scale ADF-STEM images (Fig. 3b) and their corresponding FFT pattern (Fig. 3c) confirm the precipitation of $FeTe_2$ in FGeT following the same annealing process used for FGaT. Detailed ADF-STEM image reveals the interface between the non-layered $FeTe_2$ and the FGeT layers (Fig. 3d), which is absent in pristine FGeT (Fig. S16).

The thermally annealed FGeT exhibits greater sensitivity to temperature, as evidenced by the more extensive precipitation of $FeTe_2$, with a thickness of up to 10.2 nm (Fig. S17). Moreover, a clearer ADF-STEM image (Fig. 3e) of a distorted hexagonal-shaped $FeTe_2$ along the [1 0 0] and [$\bar{1}$ 0 1] directions shows distinct high-contrast (brighter) and low-contrast (darker) spots, representing Te and Fe atom columns, respectively. The FFT pattern (Fig. 3f) further supports this interpretation. The lattice arrangement of $FeTe_2$ along the [1 0 0] direction shows Fe atoms alternating between two adjacent tellurium layers, as confirmed by the intensity line profile (Fig. 3g). However, along the [$\bar{1}$ 0 1] direction, the Fe atoms are doubly occupied between the tellurium layers (Fig. 3h), with the periodic intensity variation of Fe atoms resulting from the simultaneous presence of both [$\bar{1}$ 0 1] and [1 0 0] directions in the $FeTe_2$ structure. The ADF-STEM image of $FeTe_2$ along the [0 0 1] direction (Fig. 3i) reveals alternating undulations of Te atoms, which are corroborated by the absence of characteristic diffraction spots from the [3 0 1] $FeTe_2$ structure in the corresponding FFT pattern (Fig. 3j).

Additionally, the precipitation of $FeTe_2$ has been shown to be associated with the formation of skyrmions, as observed through MFM measurements. Fig. S18 and S19 illustrate the formation of skyrmions under an external magnetic field, both before and after annealing. In pristine FGeT, only a few isolated skyrmions are observed in the MFM image (Fig. 3k) taken at 150 K under a 0.7 kOe magnetic field. In contrast, highly-densed skyrmion lattice formed in the annealed FGeT, as evidenced by the MFM image (Fig. 3l) (150 K and 0.7 kOe). This transformation is attributed to the precipitation of FeTe$_2$ during the annealing process and resulting FeTe$_2$/FGeT heterostructure, analogous to the FeTe$_2$/FGaT heterostructure, breaks the inversion symmetry and enhances the DMI.

**Thickness-dependent FeTe$_2$ precipitation and correlated topological magnetic states**

The ability to manipulate skyrmion properties, such as size, density, and thermal stability, is pivotal for their integration into high-density memory and logic devices. However, achieving such control requires a fundamental understanding of how material parameters, particularly thickness, influence the underlying mechanisms of skyrmion formation. It has been demonstrated in recent studies that FGaT with varying thicknesses exhibit distinct skyrmions properties, including size and order[26,30,31]. Interestingly, the extant literatures appear to exhibit a thickness limit for observable skyrmions. The critical thickness limit for the formation of skyrmions may be based on this FeTe$_2$/FGaT heterostructure strategy, as illustrated in Fig. 4. Fig. 4 investigates the temperature dependence of Raman spectra in FGaT with varying thicknesses and threshold for the precipitation of FeTe$_2$. Fig. 4a and 4b display the temperature-dependent Raman spectra for 15 nm and 70 nm thick FGaT samples, respectively. At room temperature (303 K), both samples exhibit a broad peak around 150 cm$^{-1}$, characteristic of FGaT, confirming the presence of pure FGaT before thermal annealing. As the temperature increases, the thicker sample (70 nm) begins to precipitate FeTe$_2$ evidenced by a weak signal of A$_g$ mode (~120 cm$^{-1}$) of FeTe$_2$ emerging at 338 K. By 368 K, the two characteristic FeTe$_2$ peaks are clearly visible, indicating substantial precipitation. In contrast, the thinner sample (15 nm) shows subtle signal of FeTe$_2$ peaks only at 373 K, with more pronounced features appearing only at 423 K (Fig. S20). The two samples, differing in thickness, exhibit a notable disparity in the temperature at which FeTe$_2$ precipitation occurs.

To further explore the role of thickness and temperature on FeTe$_2$ precipitation,

Raman spectra were collected for samples with thicknesses ranging from 15 to 600 nm. As shown in Fig. 4c, $FeTe_2$ precipitation exhibits a strong dependence on sample thickness. The transition temperature was defined as the point at which the intensity ratio of $FeTe_2$ to FGaT Raman peaks exceeds 1/3 (Fig. S20), indicating a certain amount of $FeTe_2$ precipitation at this temperature. This approach ensures consistent quantification of $FeTe_2$ precipitation onset across thickness-dependent datasets. The transition temperature reveals a distinct threshold at approximately 25 nm. For samples thinner than 25 nm (red dots), the transition temperature ranges from 380–400 K, whereas thicker samples (orange dots) precipitate $FeTe_2$ at lower transition temperatures, between 320–350 K. This threshold indicates that thicker FGaT samples are more prone to $FeTe_2$ precipitation, with a clear distinction in the transition temperatures based on the sample thickness. Furthermore, the intensity ratio of $FeTe_2$ to FGaT Raman peaks at the highest temperature (423 K) reinforces this threshold. As shown in Fig. S20b, the ratios for thinner samples range from 1.5 to 2, while for thicker samples, the ratios increase to 3.5 to 6. This suggests that thinner FGaT samples are less likely to precipitate $FeTe_2$.

A comparison with the previous literatures (Fig. 4d) suggests that the observed thickness threshold for $FeTe_2$ precipitation closely aligns with the minimum thickness required for skyrmion formation in FGaT, as reported earlier[26,30]. This further implies that $FeTe_2$ precipitation plays a critical role in skyrmion formation by providing the necessary symmetry breaking to stabilize skyrmion lattices.

## Discussion and Outlook

In this study, we investigated the spontaneous enhanced DMI during FC in FGaT and FGeT. Through FC, we observed the creation of room-temperature-stable skyrmions in FGaT. The FC process led to the appearance of two sharp peaks in the Raman spectra, attributed to the formation of $FeTe_2$ precipitates, confirmed by phonon spectral simulations and XPS analysis.

The study revealed the spontaneous enhanced DMI in FGaT after FC, attributed to the formation of $FeTe_2$ precipitates and the subsequent development of $FeTe_2$/FGaT heterostructures. SHG spectroscopy confirmed inversion symmetry breaking in the originally centrosymmetric FGaT lattice. High-resolution STEM and EDS analyses further demonstrated $FeTe_2$ precipitation at the surfaces of FGaT, indicating a structural transformation induced by thermal decomposition. During annealing, the outermost layers of FGaT undergo decomposition, facilitated by the comparatively low bonding energy of this material. As a result, Fe and Te atoms diffuse and re-bond, leading to the nucleation of $FeTe_2$ crystallites, while excess Fe and Ga atoms migrate to the surface and remain in an amorphous state. The resulting $FeTe_2$/FGaT heterostructure introduces broken inversion symmetry, which, as observed in other van der Waals heterostructures[43-45], induces an imbalance in electronic and magnetic interactions, enhancing DMI and stabilizing skyrmion lattices. The bond-free vdW interfaces in $FeTe_2$/FGaT enable seamless integration of dissimilar materials without lattice-matching constraints, preserving spin coherence for low-power spintronic devices.

What's more, the similar phenomenon is observed in FGeT. The Raman spectroscopy and STEM analysis both demonstrate the precipitation of $FeTe_2$ and the $FeTe_2$/FGeT interface. The enhancement of DMI stabilizes skyrmions, as confirmed by MFM, which reveals a transition from labyrinthine domains in pristine samples to high-density skyrmion lattices after annealing. The field-dependent evolution of skyrmions further supports the role of $FeTe_2$ in tuning the balance between PMA and interfacial DMI. This cross-family consistency underscores the critical role of $FeTe_2$ in breaking inversion symmetry and mediating spin-orbit interactions at heterointerfaces. Such insights could guide exploration of tailored DMI in other emerging vdW ternary Tellurides through controlled phase separation.

Additionally, the thickness of FGaT plays a critical role in determining the precipitation temperature of $FeTe_2$ and the formation of skyrmions. Thicker samples exhibit $FeTe_2$ precipitation at lower temperatures compared to thinner ones, revealing a strong thickness dependence in the phase transformation process. A critical thickness threshold of ~25 nm was identified, below which $FeTe_2$ precipitation was significantly suppressed. This threshold also correlated with the emergence of stable skyrmions, suggesting that FeTe$_2$ precipitation is a key factor in enabling skyrmion formation. The critical thickness (~25 nm) links FeTe$_2$ precipitation to skyrmion stability, offering a tunable knob for device optimization. Thickness gradients or vertical stacking could spatially engineer skyrmion size and mobility, enabling programmable spin textures in scalable vdW platforms.

**Summary**

In summary, our findings underscore the significance of FeTe$_2$ precipitation in the enhancing DMI and stabilizing skyrmions in FGaT and FGeT. The study provides new insights into the mechanisms behind the origin of the DMI in ternary tellurides. The ability to control skyrmion formation through thermal treatment provides a new avenue for designing magnetically tunable materials, where phase engineering can be leveraged to tailor DMI strength and skyrmion stability, contributing to the advancement of spintronic applications.

# Methods

**Fe₃GaTe₂ single crystal growth**

FGaT crystals were grown using the self-flux method. Highly pure Fe (99.9 %), Ga (99.99 %), and Te (99.99 %) powders, with a mole ratio of 1:1:2, were mixed under Ar atmosphere in a glove box and sealed in a quartz tube under 1.9 Pa. To capture the excess flux during centrifugation, a crucible filled with quartz wool was placed on top of the growth crucible. The melt was homogenized at 1000 °C for approximately 24 hours, then cooled quickly to 880 °C in 1 hour, followed by slow cooling to 780 °C in 100 hours. At this temperature, the ampoules were removed from the furnace and placed in a centrifuge to expel the excess flux. Then large single crystals could be obtained. FGaT nanoflakes were mechanically exfoliated onto a freshly cleaned SiO$_2$/Si substrate. In our experiment, the pristine FGaT nanoflakes were mechanically exfoliated onto a freshly cleaned SiO$_2$/Si substrate with the thicknesses ranging from 12 to 600 nm (Fig. S1).

**Magnetic force microscopy measurements**

MFM measurements on FGaT were performed using the DIMENSION ICON (Bruker). The magnetic field during FC was generated by a neodymium magnet. The measurement was conducted in a glove box protected by high-purity nitrogen gas at room temperature. The magnetic tip from Nanoworld coated with Co/Cr was used for all measurements. The force constant and resonance frequency of this tip are 3 N/m and 75 kHz, respectively. During the measurement, the topography of the sample was first acquired in tapping mode. During the second scan, the MFM tip was lifted by 100 nm above the sample to measure the magnetic signal in tapping mode. MFM measurements on FGeT were carried out using variable temperature MFM system (attocube-2100). Hard magnetic point probes with cobalt coating (Nanosensors) were employed. The samples were cooled from room temperature to 150 K in a low-pressure helium gas environment under zero magnetic field. Prior to MFM imaging, a relatively flat region was identified using tapping mode AFM, and slope compensation was performed to ensure a constant tip-sample distance across the scan area. Phase images were then acquired at a fixed tip-sample separation, typically set to 100 nm.

**Raman Spectroscopy measurements**

The Raman measurements were obtained by a Witec Alpha 300R confocal Raman microscope. The 532 nm linear polarized laser was focused perpendicular to the surface of the sample through a 100x objective lens (numerical aperture = 0.9). The maximum power is kept below 0.2 mW to avoid the laser thermal effects on the sample. The Raman signals were first collected by a photonic crystal fiber and then coupled into the spectrometer with 1800 g/mm grating.

**HAADF-STEM**

STEM measurements were performed using a Cs-corrected FEI Themis, equipped with a cold field-emission gun operating at 200 kV. The resolution ADF-STEM images were collected with a convergence semiangle of 24.5 mrad, collecting scattering from >44 mrad. Image filtered using Gaussian functions.

**Theoretical calculation**

The relaxed crystal structures were obtained using the Vienna Ab initio Simulation Package (VASP) within the framework of density functional theory (DFT). The exchange-correlation functional was described using the generalized gradient approximation (GGA) as parameterized by Perdew, Burke, and Ernzerhof (PBE). The plane-wave energy cutoff and convergence threshold were set to 450 eV and $10^{-6}$ eV, respectively. Phonon spectra with a $2\times2\times2$ supercell were calculated using the finite-displacement method implemented in the Phonopy code.

# Data availability

Source data are provided with this paper. All other data that support the findings of this study are available from the corresponding authors upon reasonable request.

# Competing interests

The authors declare no competing interests.

# Acknowledgement

We acknowledge the assistance from the National Key Research and Development

Program of China (No.2021YFA 1200800 and No.2022YFB3403400), the National Natural Science Foundation of China (Grant Nos. 12274067, 92464101, T2321002 and 12274016) and the Start-up Funds of Wuhan University.

## Author Contributions

S.X., L.Y. and I.A.M. contributed equally to this work. S.X. designed the experiments with supervision from S.L. and T.Y. H.S. grew the crystal with supervision from Y.D. S.X. fabricated the samples and performed the optical measurements with help from Y.S. and H.L. Y.L. conducted STEM characterization and data analysis with supervision from C.Z. I.A.M. and X.Z. performed the MFM and AFM measurements. K.H. performed theoretical simulations with supervision from Q.C. S.L., C.Z. and T.Y. supervised the project. S.X. and Y.L. co-wrote the manuscript in consultation with all the authors. All the authors discussed the results.

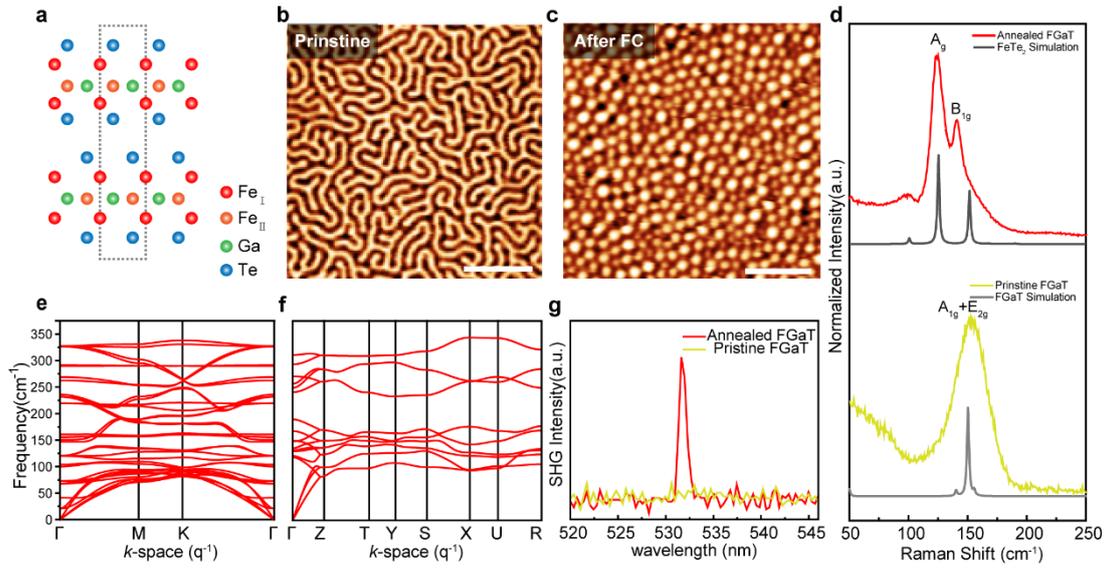

**Figure 1 | Creation of robust skyrmions and emergence of anomalous Raman signatures in Fe₃GaTe₂ (FGaT) after field cooling (FC)**. **(a)** Side view of the crystal structure of FGaT, illustrating the atomic arrangement of Fe, Ga, and Te atoms. **(b, c)** Magnetic force microscopy (MFM) images of FGaT in its pristine state **(b)** and after FC **(c)**, showing a transition from a labyrinth-like domain structure to skyrmions state. **(d)** Raman spectra of pristine and annealed FGaT, highlighting distinct vibrational modes and anomalous spectral changes upon annealing. **(e, f)** Calculated phonon dispersion relations for FGaT and FeTe₂, revealing vibrational characteristics. **(g)** Second harmonic generation (SHG) spectra of pristine and annealed FGaT, demonstrating symmetry breaking. Scale bars in **b** & **c** is 1 μm.

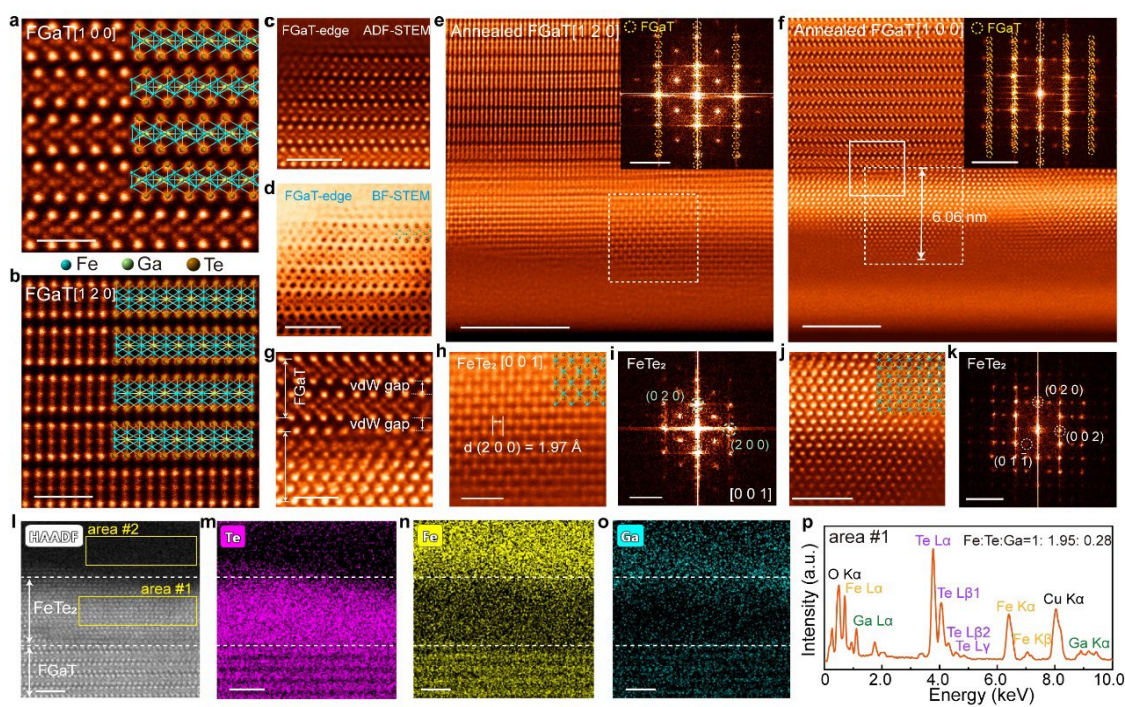

**Figure 2 | Formation of FeTe$_2$/FGaT interface after the annealing of FGaT. (a, b)** Atomic-resolution annular dark-field scanning transmission electron microscopy (ADF-STEM) images of pristine FGaT along [1 0 0] **(a)** and [1 2 0] **(b)** zone axes, showing the atomic arrangement. **(c, d)** ADF-STEM **(c)** corresponding bright-field STEM (BF-STEM) **(d)** image of in the pristine FGaT edge. **(e, f)** Large-scale ADF-STEM images of annealed FGaT along [1 2 0] **(e)** and [1 0 0] **(f)** zone axes, with insets displaying the corresponding fast Fourier transform (FFT) patterns. **(g)** Magnified ADF-STEM image of the white solid boxed region in **(f)**, highlighting structural changes upon annealing. **(h, j)** Enlarged ADF-STEM images of FeTe$_2$ extracted from the white dashed regions in **(e)** and **(f)**, respectively. **(i, k)** Corresponding FFT patterns for **(h)** and **(j)**, confirming the crystalline structure of FeTe$_2$. **(l)** Cross-sectional STEM image of FGaT/FeTe$_2$ heterostructure with energy-dispersive X-ray spectroscopy (EDS) elemental mapping for **(m)** Te, **(n)** Fe, and **(o)** Ga elements. **(p)** Quantitative EDS spectrum analysis from area 1 in **(l)** indicating an Fe:Te ratio of 1:2, confirming the precipitation of FeTe$_2$. Scale bars: 1 nm in **(a, b, g, h)**; 2 nm in **(c, d, j, l-o)**, **(j)** and **(l-o)**; 5 nm in **(e, f)**; 5 nm$^{-1}$ in FFT inset of **(e, f, i, k)**

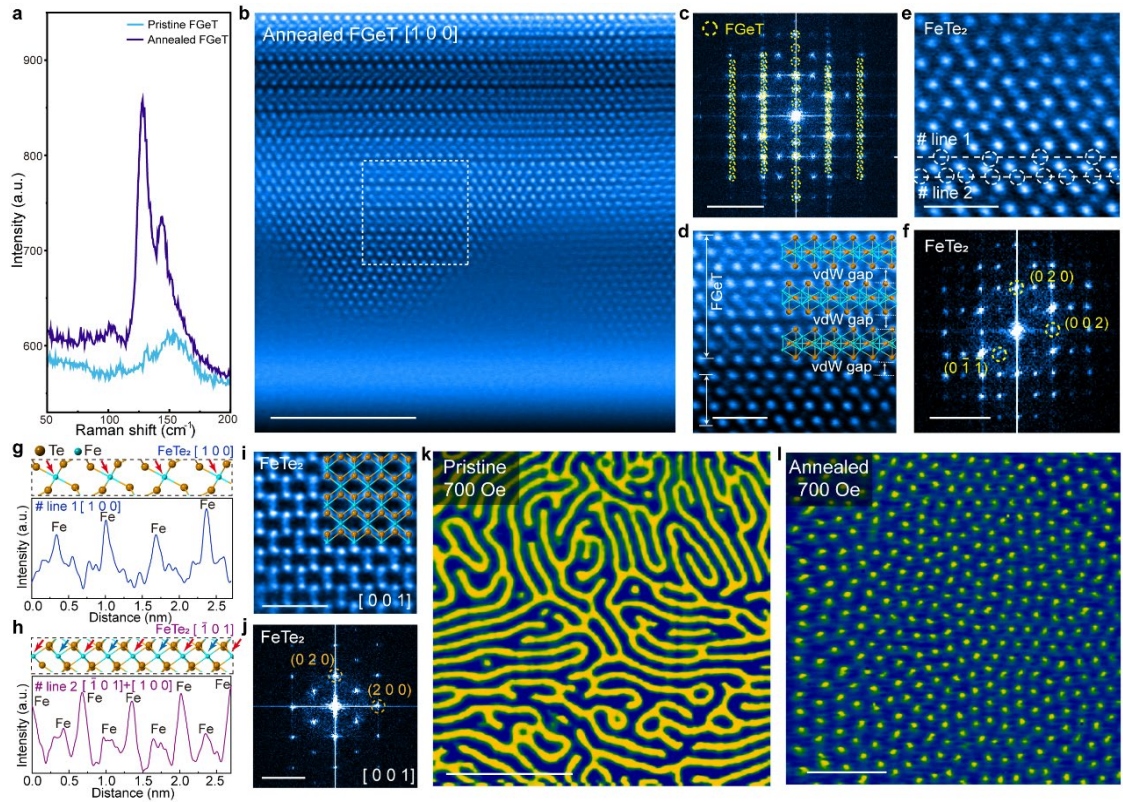

**Figure 3 | Enhanced Dzyaloshinskii-Moriya interaction (DMI) in Fe₃GeTe₂ (FGeT).**
(a) Raman spectra of pristine and annealed FGeT. (b) Large-scale ADF-STEM image of annealed FGeT along [1 0 0] zone axis. (c) Corresponding fast Fourier transform (FFT) pattern pattern of **b**, showing the crystallographic features. (d) Magnified ADF-STEM image from the white boxed region in **b**. (e) Atomic-resolution ADF-STEM image of FeTe₂ viewed along the [1 0 0] direction of annealed FGeT, with (f) the corresponding FFT pattern. (g, h) Intensity line profiles along white dashed line 1 and line 2 in **f**, highlighting structural differences of FeTe₂ along the [1 0 0] and [$\bar{1}$ 0 1] directions. (i) Atomic-resolution ADF-STEM image of FeTe₂ viewed along the [1 2 0] direction of annealed FGeT, with (j) the corresponding FFT pattern. (k) MFM image of FGeT before annealing under a magnetic field of 0.7 kOe at 150K, showing only a few isolated skyrmions. (l) MFM image of annealed FGeT under a magnetic field of -0.7 kOe at 150K, revealing a high-density skyrmions. Scale bars: 5 nm (**b**); 1 nm (**d-e, i**); 5 nm$^{-1}$ in (**c, f, j**); 5 μm (**k, l**).

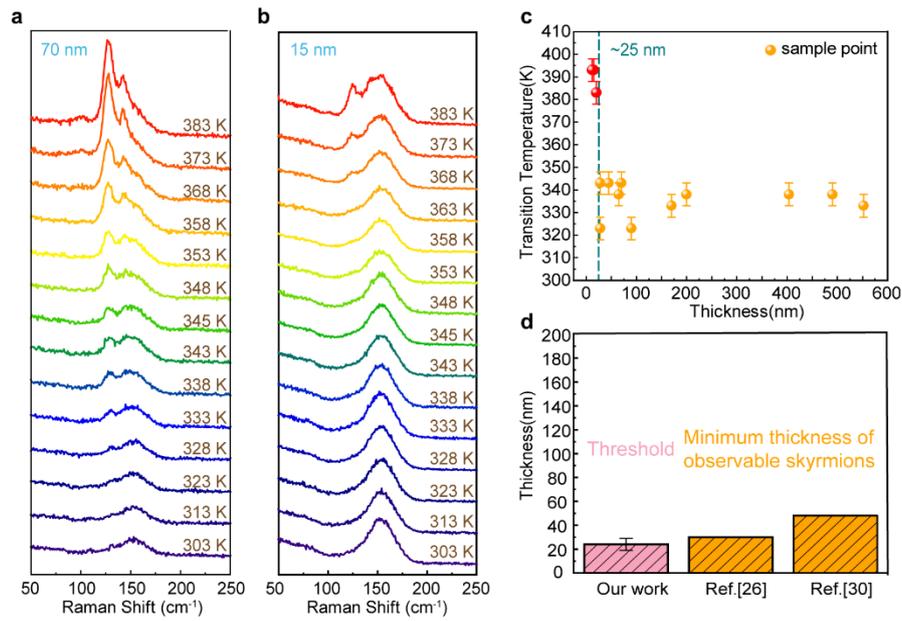

**Figure 4 | Thickness and temperature dependent Raman of FGaT**. **(a, b)** Temperature-dependence Raman spectra of FGaT with thicknesses of 70 nm **(a)** and 15 nm **(b)** **(c)** Thickness dependence of the transition temperature for the formation of FeTe$_2$, indicating a threshold around 25nm. **(d)** Comparison of threshold thickness for forming FeTe$_2$ and creating skyrmions, based on previous reports and this work.

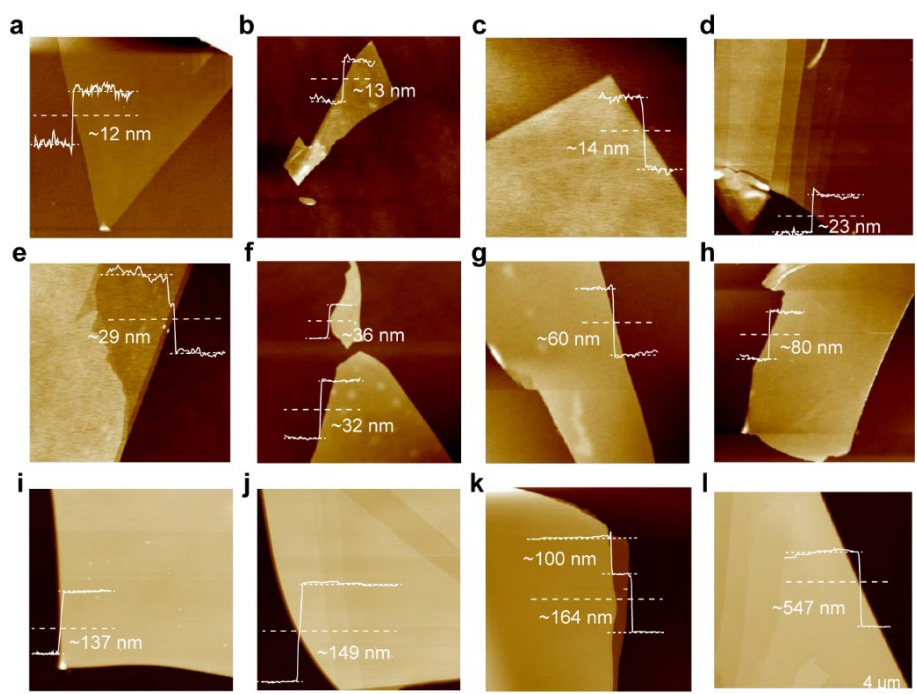

**Figure S1 | Atomic force microscopy (AFM) characterization of FGaT flakes with varying thicknesses.** Atomic force microscopy (AFM) images of FGaT flakes with varying thicknesses. **(a–l)** AFM topography scans of exfoliated FGaT flakes, illustrating thickness measurements ranging from ~12 nm to ~547 nm, as determined via step height analysis. White dashed lines denote the locations of the height profiles, with corresponding step height plots provided in each panel.

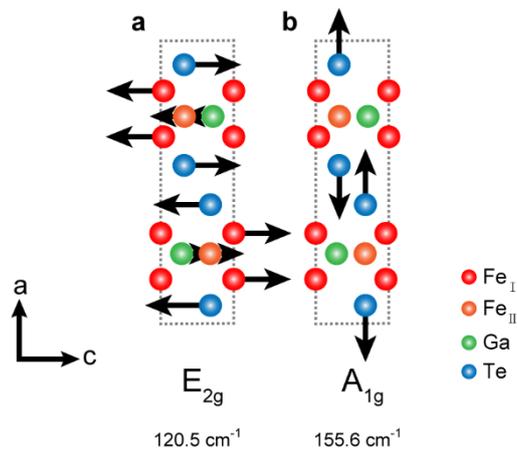

**Figure S2 | Calculated atomic displacement patterns of Raman-active modes in bulk FGaT. (a)** $E_{2g}$ mode at 120.5 cm$^{-1}$, corresponding to the in-plane vibrational motion of Fe, Ga, and Te atoms. **(b)** $A_{1g}$ mode at 155.6 cm$^{-1}$, representing out-of-plane vibrations. Black arrows denote the direction of atomic displacements. These vibrational modes serve as characteristic signatures of FGaT, providing key insights into lattice dynamics and phase transformations.

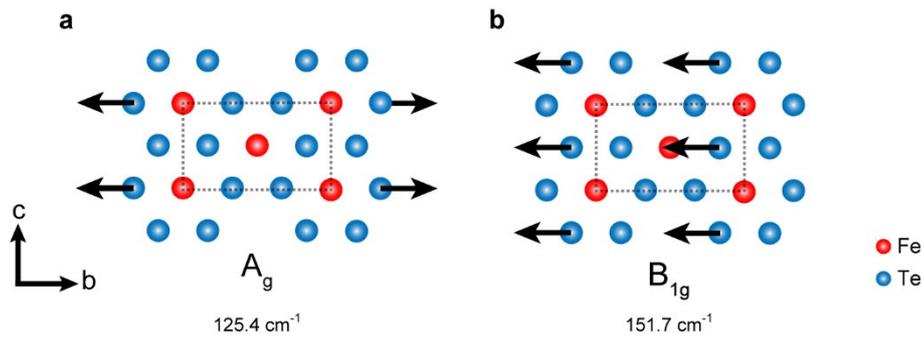

**Figure S3 | Calculated atomic displacement patterns of Raman-active modes in bulk FeTe$_2$. (a)** $A_g$ mode at 125.4 cm$^{-1}$, corresponding to symmetric in-plane vibrations of Fe and Te atoms. **(b)** $B_{1g}$ mode at 151.7 cm$^{-1}$, associated with asymmetric atomic displacements. Black arrows denote the direction of atomic motion. These vibrational modes serve as characteristic signatures of FeTe$_2$, offering insight into its lattice dynamics and structural properties.

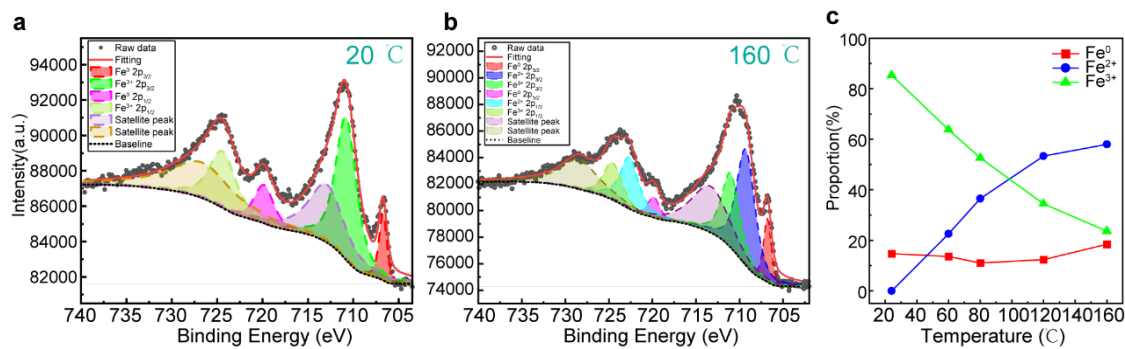

**Figure S4 | X-ray photoelectron spectroscopy (XPS) analysis of the exfoliated fresh FGaT crystal surface. (a, b)** High-resolution Fe 2p spectra recorded at 20°C **(a)** and 160°C **(b)**, illustrating the evolution of Fe valance states. Spectral deconvolution reveals contributions from $Fe^0$, $Fe^{2+}$, and $Fe^{3+}$, indicating temperature-induced modifications in the chemical environment. **(c)** Temperature-dependent evolution of Fe valence states, showing a progressive increase in $Fe^{2+}$ content with temperature, consistent with $FeTe_2$ precipitation, while $Fe^{3+}$ concentration decreases. These findings support the temperature-driven transformation of FGaT and the emergence of $FeTe_2$.

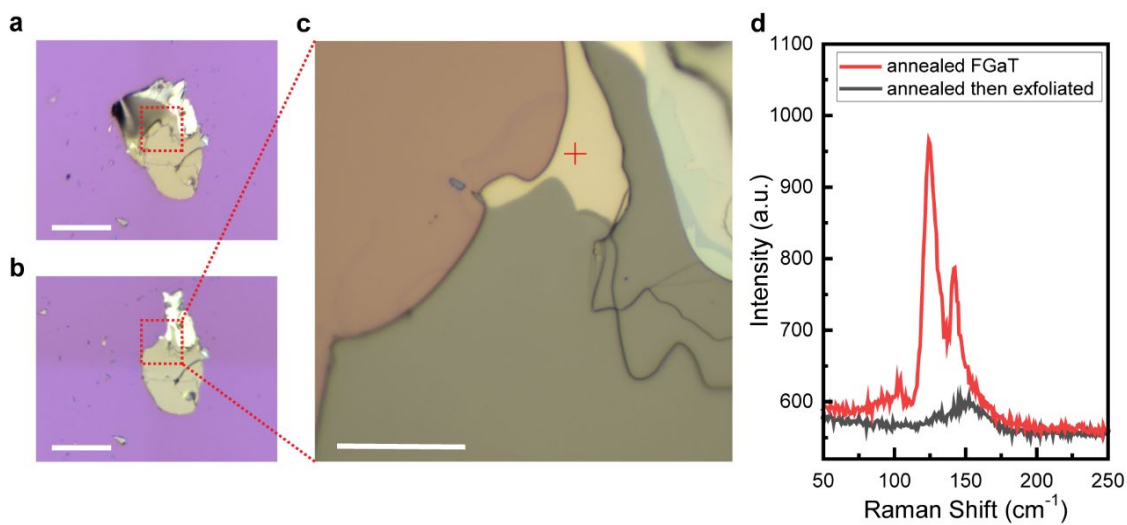

**Figure S5 | Raman analysis of re-exfoliated annealed FGaT (a)** Optical image of annealed FGaT prior to exfoliation. **(b, c)** Optical images of FGaT after re-exfoliation, revealing the exposed inner layers. **(d)** Raman spectra comparison between annealed FGaT (red) and the re-exfoliated layer (black). The re-exfoliated sample displays only the characteristic peaks of FGaT, while $FeTe_2$-related peaks are absent, confirming that $FeTe_2$ precipitates predominantly on the surface following annealing. Scale bars: 100 μm **(a, b)**, 20 μm **(c)**.

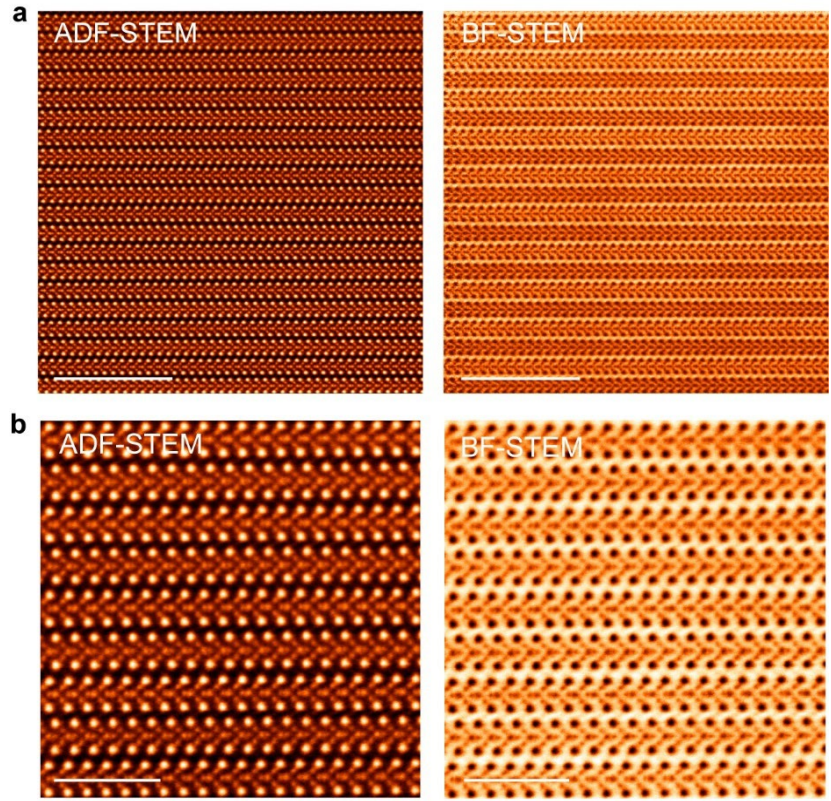

**Figure S6 | The atomic structure of pristine FGaT. (a, b)** Annular dark field **(**ADF)-STEM images (left panel) and corresponding BF (bright field)-STEM images (right panel) of pristine FGaT taken along [1 0 0] zone axis. Scale bars: 2 nm in **a**; 5 nm in **b**.

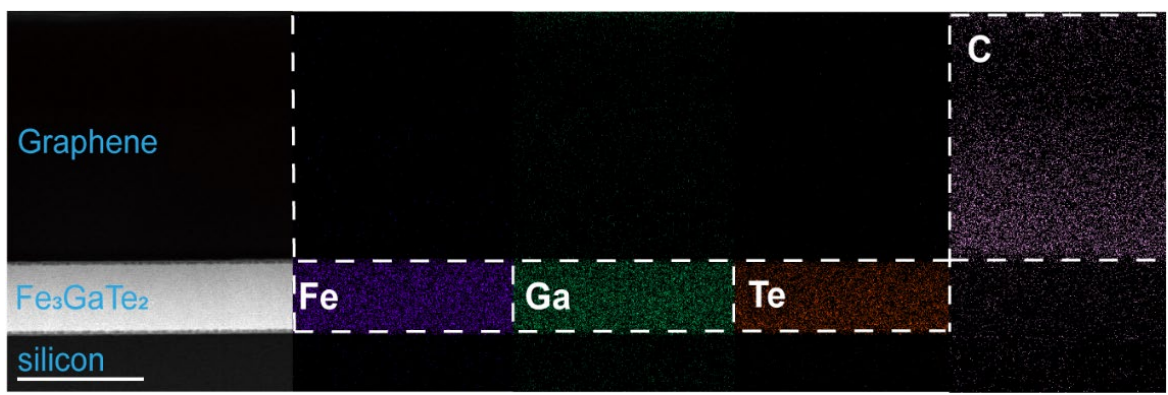

**Figure S7 | Corres-sectional ADF-STEM image of Graphene capped-FGaT and corresponding Energy dispersive X-ray spectroscopy (EDS) mapping.**

STEM image of the FGaT with a graphene capping layer and corresponding EDS elemental mapping. The EDS maps confirm the spatial distribution of Fe (purple), Ga (green), Te (orange), and C (pink), illustrating the composition of the FGaT layer, graphene capping, and underlying silicon substrate. Scale bars: 200 nm.

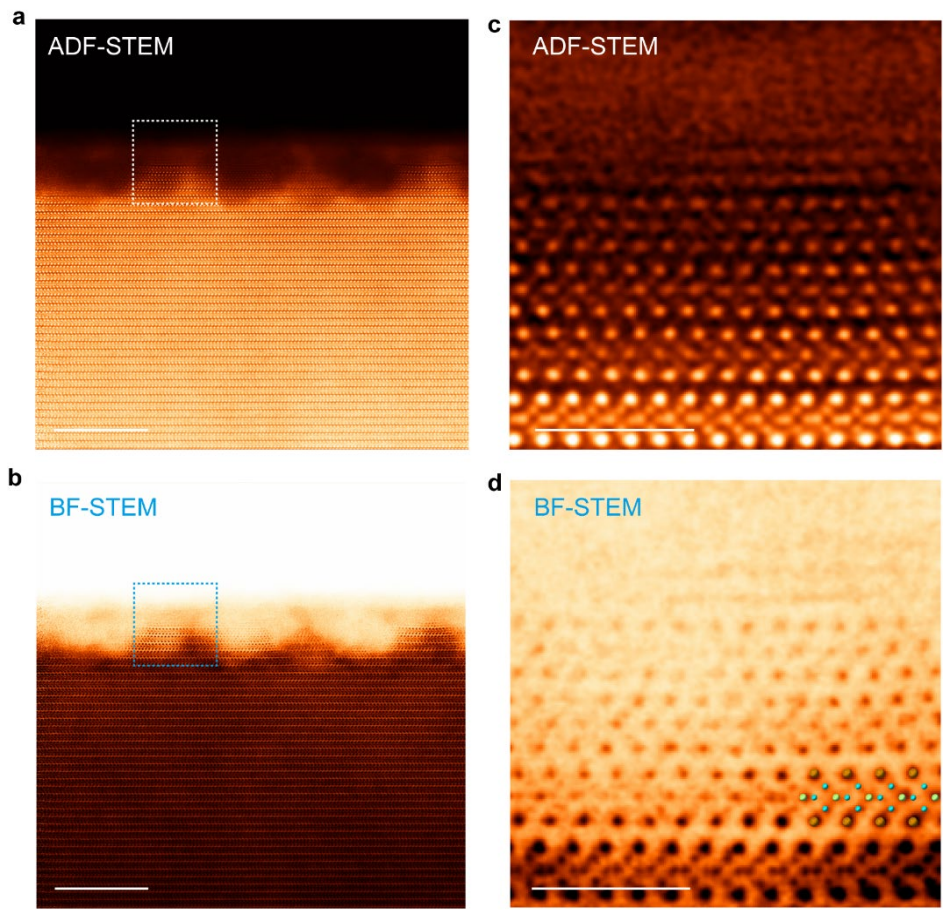

**Figure S8 | HAADF-STEM images of the edge of pristine FGaT. (a)** ADF-STEM image and **(b)** corresponding BF-STEM image showing the edge of pristine FGaT. **(c, d)** zoomed in STEM images of the dotted squares from **a** and **b** respectively. The color balls in **d** represent Fe (blue), Te (orange), and Ga (green) atoms, respectively. Scale bars: 10 nm in **a-b**; 2 nm in **c-d**.

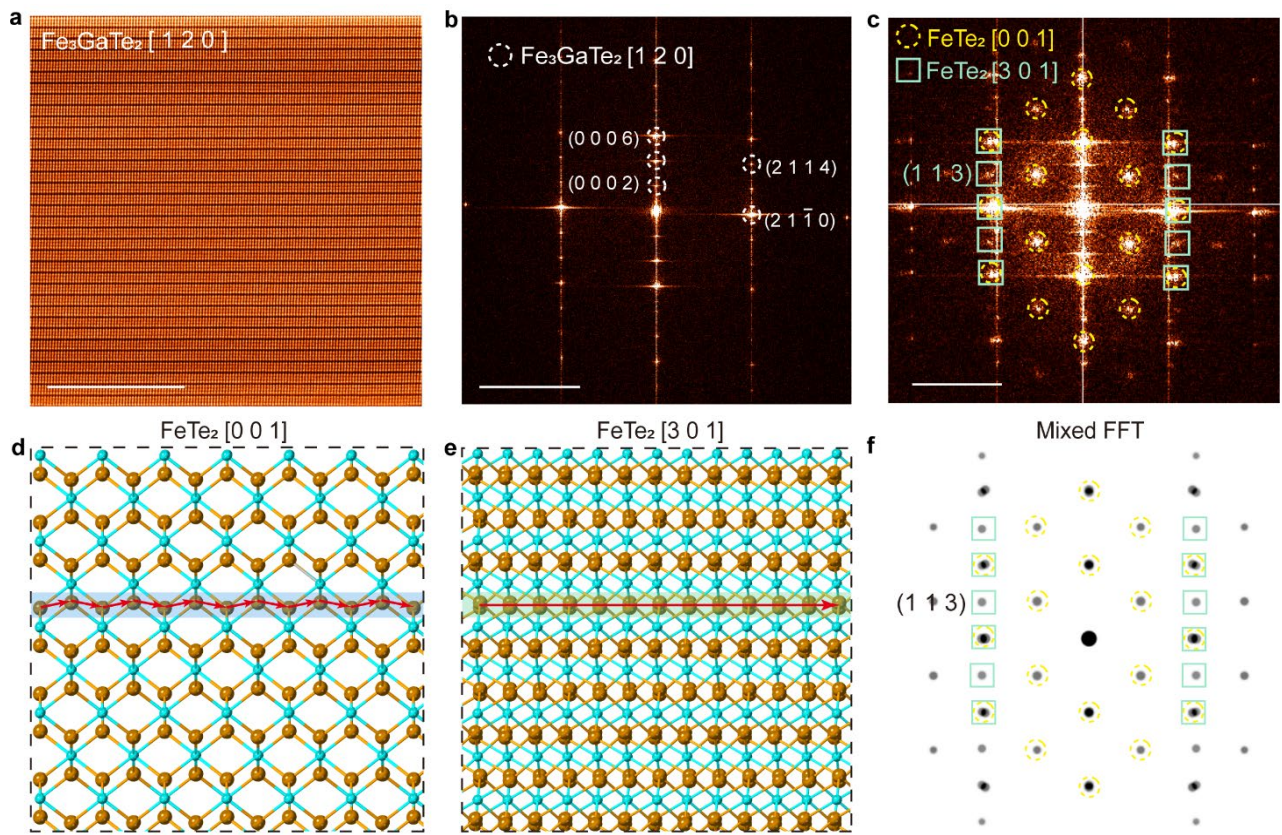

**Figure S9 | Atomic structure of FeTe₂ formed along the [1 2 0] direction in annealed FGaT.** (a) ADF-STEM image of pristine FGaT along the [1 2 0] zone axis. (b) Corresponding fast Fourier transform (FFT) image of pristine FGaT, showing its characteristic diffraction spots. (c) FFT pattern of annealed FGaT from Figure 2e, highlighting diffraction spots associated with FeTe₂ [0 0 1] (yellow circles) and FeTe₂ [3 0 1] (green squares), confirming the coexistence of FeTe₂ in multiple orientations. (d, e) Atomic models of FeTe₂ viewed along the [0 0 1] and [3 0 1] zone axes, respectively, illustrating structural variations, with red arrows indicating Te atomic displacement differences between the two orientations. (f) Simulated mixed FFT pattern incorporating FeTe₂ [0 0 1] and [3 0 1] diffraction spots, demonstrating excellent agreement with the experimental FFT pattern in (c). Scale bars: 10 nm in **a**, **d**; 5 nm$^{-1}$ in **b** and **c**.

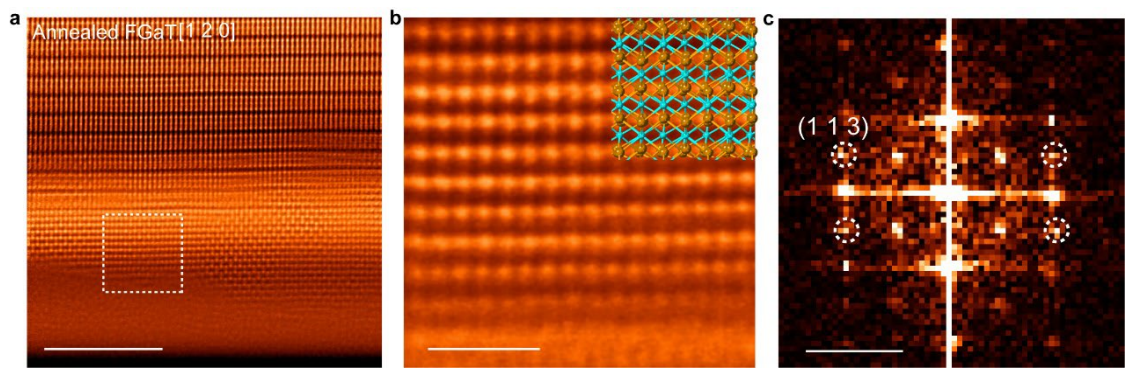

**Figure S10 | The formation of [3 0 1] FeTe$_2$ along the [1 2 0] direction of annealed FGaT. (a)** Large scale ADF-STEM image of annealed FGaT along the [1 2 0] zone axis. **(b)** High-resolution ADF-STEM image stacking of [0 0 1] and [3 0 1] FeTe$_2$, derived from white dashed square **(a)**. **(c)** Corresponding FFT image showing characteristic (1 1 3) diffraction spots. The observed spots from [0 0 1] FeTe$_2$ are likely attributed to the stacking of [0 0 1] and [3 0 1] FeTe$_2$. Scale bars: 5 nm in **a**; 1 nm in **b**; 5 nm$^{-1}$ in **c.**

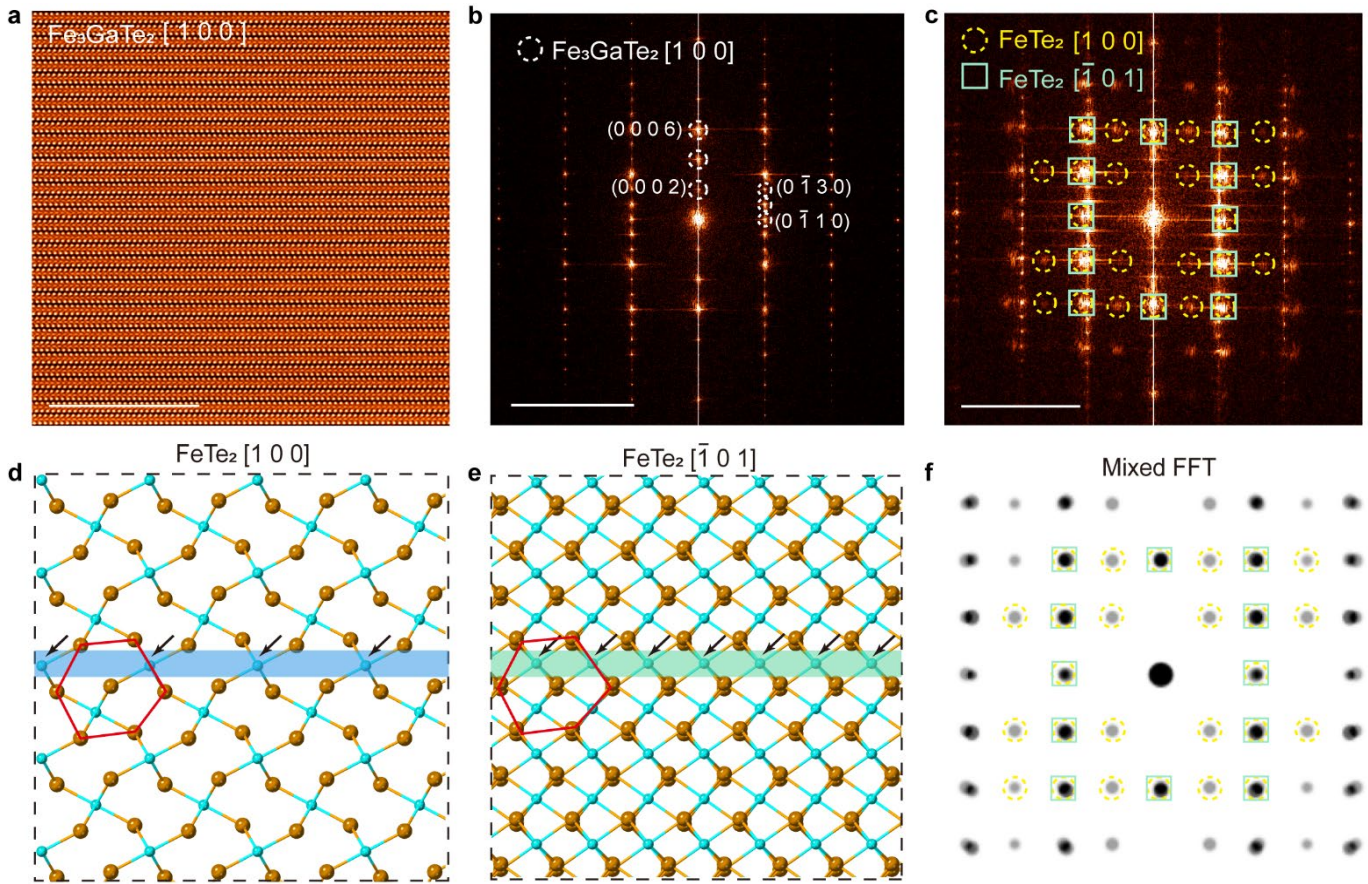

**Figure S11 | Atomic structure of FeTe$_2$ formed along the [1 0 0] direction in annealed FGaT.** **(a)** ADF-STEM image of pristine FGaT taken along [1 0 0] zone axis and **(b)** Corresponding fast Fourier transform (FFT) image of pristine FGaT, showing its characteristic diffraction spots. **(c)** FFT pattern of annealed FGaT from Figure 2f, highlighting diffraction spots associated with FeTe$_2$ [1 0 0] (yellow circles) and FeTe$_2$ [$\bar{1}$ 0 1] (green squares), confirming the coexistence of FeTe$_2$ in multiple orientations. **(d, e)** Atomic models of FeTe$_2$ viewed along the [1 0 0] and [$\bar{1}$ 0 1] zone axes, respectively, illustrating structural variations, with red arrows indicating Te atomic displacement differences between the two orientations. **(f)** Simulated mixed FFT pattern incorporating FeTe$_2$ [1 0 0] and [$\bar{1}$ 0 1] diffraction spots, demonstrating excellent agreement with the experimental FFT pattern in **(c)**. Scale bars: 10 nm in **a**, **d**; 5 nm$^{-1}$ in **b** and **c**.

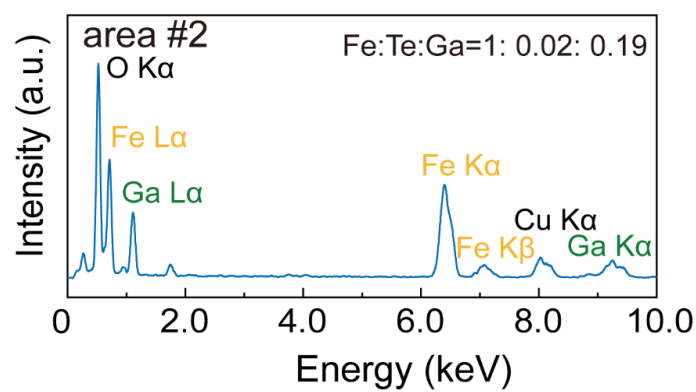

**Figure S12 | EDS spectrum of Area #2 from Figure 2l, revealing the elemental composition of the FGaT layer above the FeTe$_2$ region.**

EDS spectrum of Area #2 from Fig. 2l, showing the elemental distribution within the FGaT layer above the FeTe$_2$ region. The measured Fe: Te: Ga ratio of 1: 0.02: 0.19 indicates significant Te depletion, suggesting that this region consists primarily of Fe rather than FeTe$_2$. These findings indicate that Area #2 corresponds to an Fe-rich phase, potentially amorphous or disordered Fe, resulting from the excess Fe during annealing.

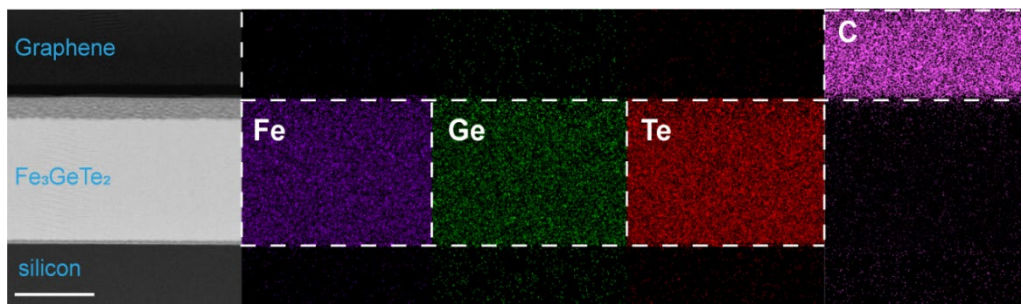

**Figure S13 | Corres-sectional ADF-STEM image of Graphene capped-FGeT and corresponding Energy dispersive X-ray spectroscopy (EDS) mapping.**

STEM image of the FGeT with a graphene capping layer and corresponding EDS elemental mapping. The EDS maps confirm the spatial distribution of Fe (purple), Ge (green), Te (orange), and C (pink), illustrating the composition of the FGeT layer, graphene capping, and underlying silicon substrate. Scale bars: 200 nm.

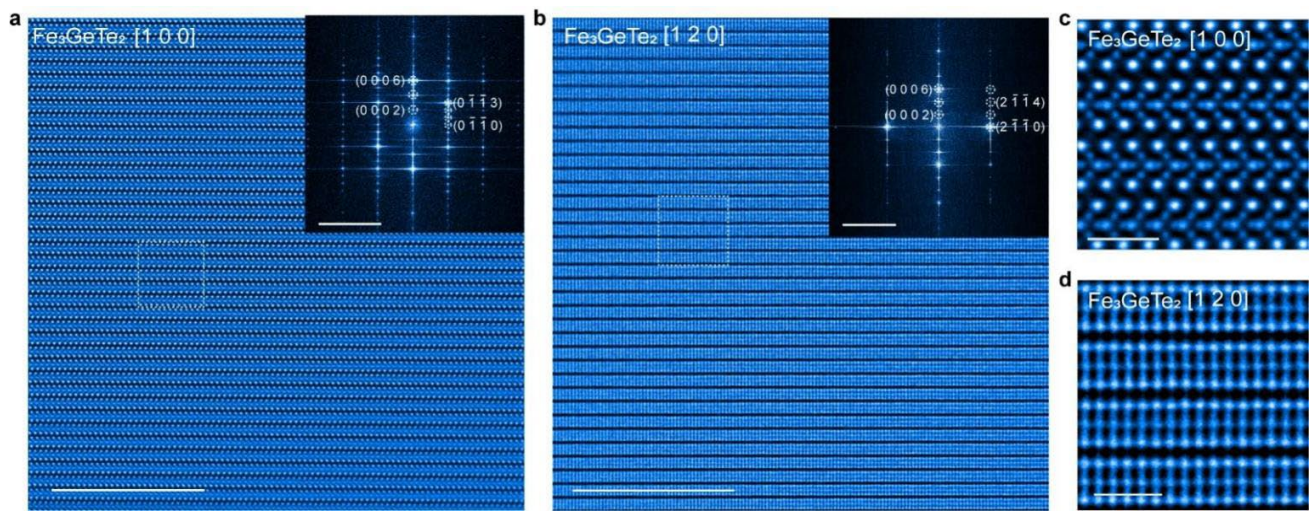

**Figure S14 | HAADF-STEM images of atomic structure of pristine FGeT. (a, b)** ADF-STEM images of pristine FGeT viewed along [1 0 0] and [1 2 0] zone axes, respectively. Insets display the corresponding FFT patterns, confirming the crystallographic orientation. **(c, d)** High-resolution ADF-STEM images of white dashed squares in **(a)** and **(b)**, respectively, showing detailed atomic arrangements. Scale bars: 10 nm in **a-b**; 5 nm$^{-1}$ in inset of **a** and **b**; 1 nm in **c-d**.

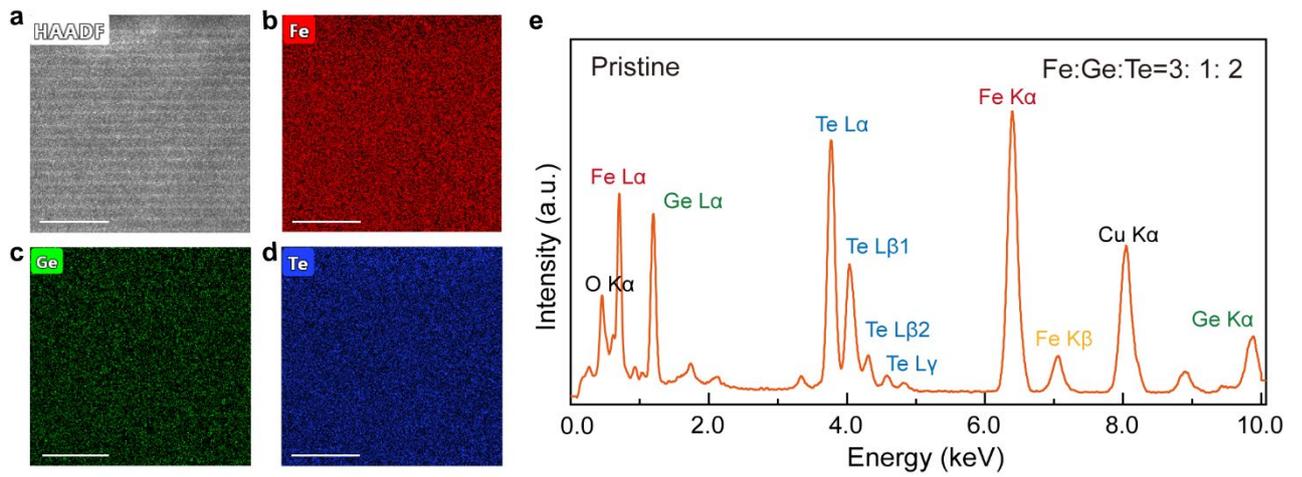

**Figure S15 | The EDS mapping and spectrum of pristine FGeT. (a)** Cross-sectional STEM images of pristine FGeT, along with corresponding EDS elemental mappings for **(b)** Fe, **(c)** Ge and **(d)** Te elements. **(e)** Quantitative analysis of EDS revealing the ratios of Fe: Ge: Te is 3: 2: 1. Scale bars: 5 nm.

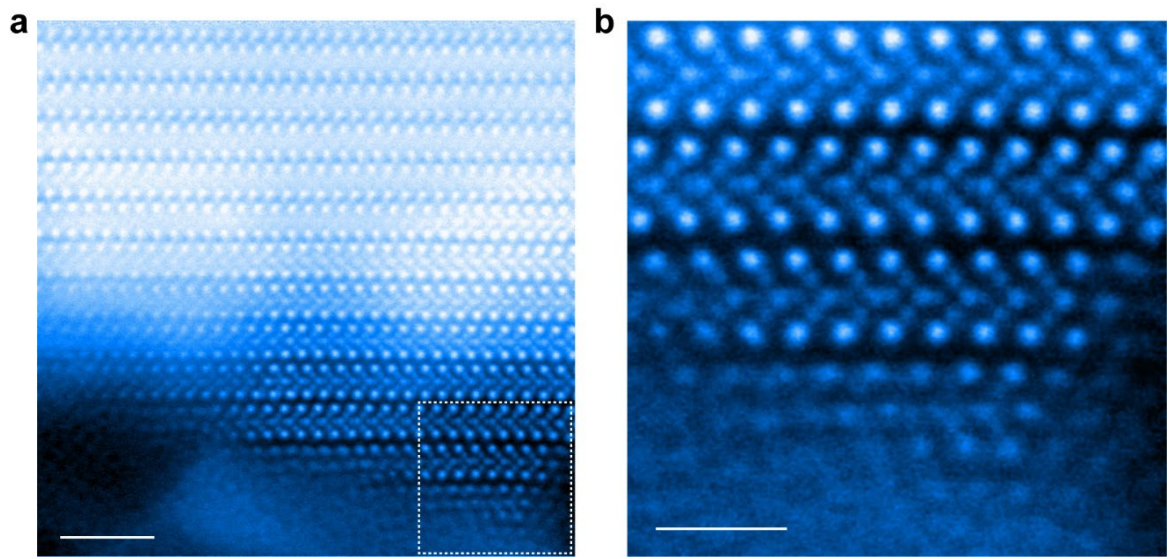

**Figure S16 | HAADF-STEM images of the edge of pristine FGeT. (a)** ADF-STEM image showing the edge of pristine FGeT. **(b)** zoomed in STEM images of the dotted squares from **a**. Scale bars: 10 nm in **a-b**; 2 nm in **c-d**.

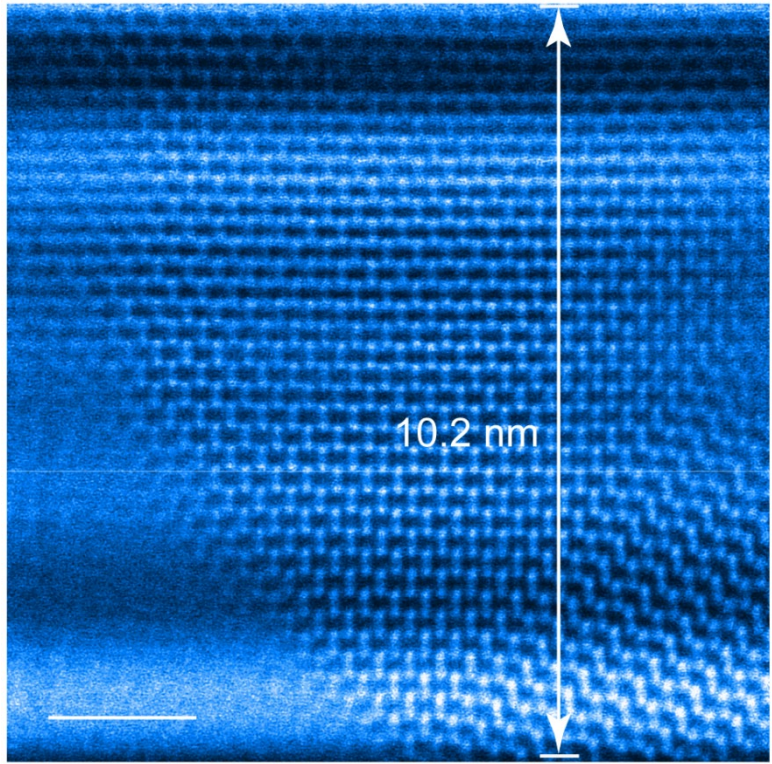

**Figure S17 | ADF-STEM image of large-scale FeTe$_2$ after annealed FGeT.** Scale bar: 2 nm.

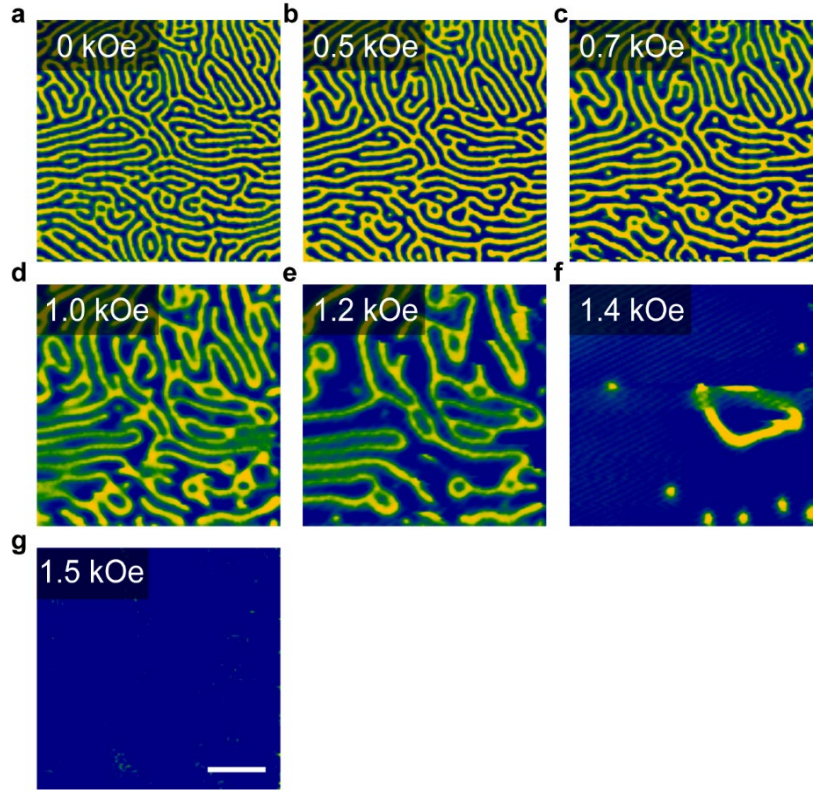

**Figure S18 | Field-dependent magnetic force microscopy (MFM) imaging of pristine FGeT. (a–g)** MFM images obtained under increasing external magnetic fields from 0 kOe to 1.5 kOe, illustrating the evolution of magnetic domain structures. At 0 kOe **(a)**, a well-defined labyrinth-like domain pattern is observed. As the field increases to 0.5 kOe **(b)** and 0.7 kOe **(c)**, the labyrinthine structure remains stable. At 1.0 kOe **(d)** and 1.2 kOe **(e)**, domains begin to fragment, giving rise to isolated skyrmions. At 1.4 kOe **(f)**, skyrmions become more pronounced, forming bubble-like magnetic textures. By 1.5 kOe **(g)**, the contrast vanishes, indicating that the system has reached a fully magnetized state with domains fully aligned along the applied field. The scale bar is 5 μm.

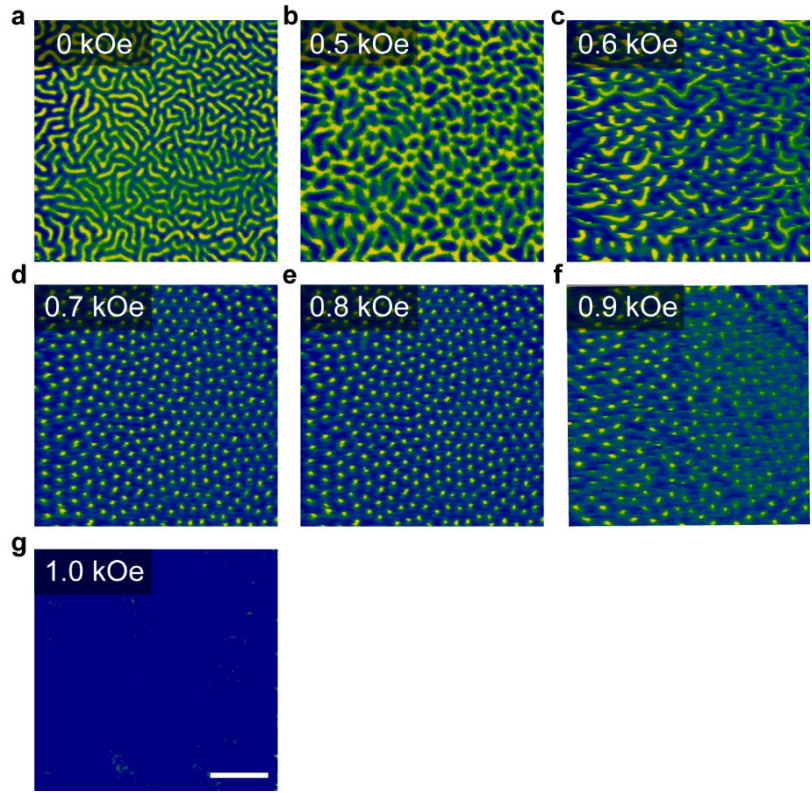

**Figure S19 | Field-dependent magnetic force microscopy (MFM) imaging of annealed FGeT. (a–g)** MFM images obtained under increasing external magnetic fields from 0 kOe to 1.0 kOe, illustrating the evolution of magnetic domain structures. At 0 kOe **(a)**, a well-defined labyrinth-like domain pattern is observed. As the field increases to 0.5 kOe **(b)** and 0.6 kOe **(c)**, domains begin to fragment, giving rise to skyrmion lattices. At 0.7 kOe **(d)** and 0.8 kOe **(e)**, a high-density skyrmion lattice is observed. At 0.9 kOe **(f)**, skyrmion lattice remains to be stable, but the contrast becomes vague. By 1.0 kOe **(g)**, the contrast vanishes, indicating that the system has reached a fully magnetized state with domains fully aligned along the applied field. The scale bar is 5 μm.

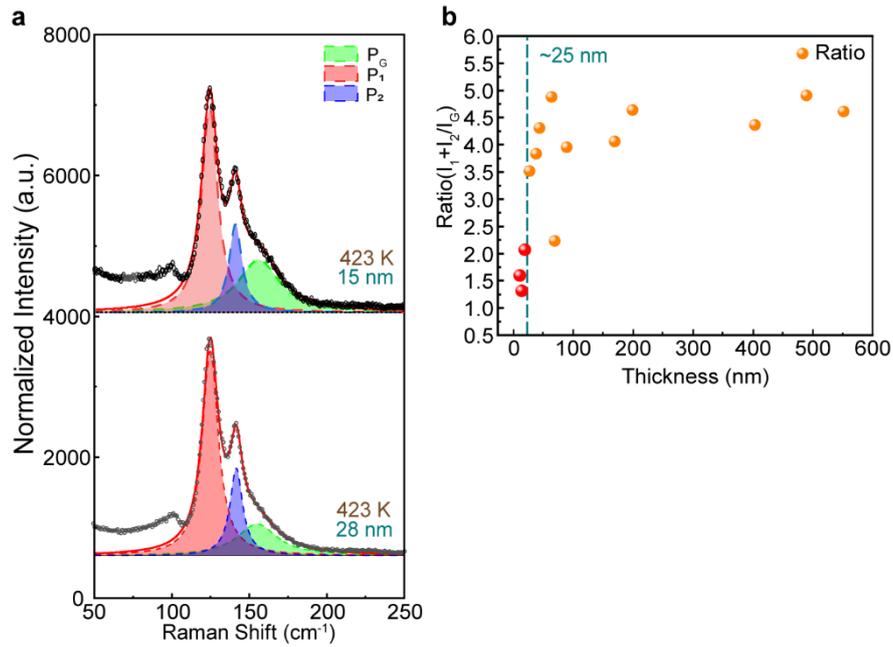

**Figure S20 | Thickness and temperature dependence of Raman spectra in FGaT (a)** 15 nm (top) and 28 nm (bottom) FGaT at 423 K, illustrating the decomposition of spectral contributions. $P_G$ is fitted Raman peak of FGaT, while $P_1$ and $P_2$ correspond to Raman peaks associated with $FeTe_2$. The presence of $FeTe_2$ peaks confirms the temperature-driven phase transformation.**(b)**Thickness dependence of the intensity ratio of $(I_{FeTe_2}/I_{FGaT})$ at 423K, revealing a critical thickness threshold of approximately 25 nm. Samples below 25 nm (red dots) exhibit a lower $FeTe_2$ intensity ratio, indicating reduced $FeTe_2$ formation, whereas samples exceeding 25 nm (orange dots) show a pronounced increase in $FeTe_2$ signal, suggesting enhanced precipitation over the thickness thershold.